\documentclass{article}
\usepackage[a4paper]{geometry}

\usepackage[usenames,dvipsnames,svgnames,table]{xcolor} 
\usepackage{microtype}
\usepackage{amsmath}
\usepackage{amsfonts}
\usepackage{comment}
\usepackage{bm}
\usepackage[numbers,sort&compress]{natbib}
\usepackage{enumitem}
\usepackage{mathtools}
\usepackage{tikz}
\usepackage{subfigure}
\usetikzlibrary{arrows,decorations.pathmorphing,backgrounds,fit,positioning,shapes.symbols,chains}
\usepackage{authblk}

\usepackage{hyperref}
\ifdefined\nohyperref\else\ifdefined\hypersetup
  \definecolor{mydarkblue}{rgb}{0,0.08,0.45}
  \definecolor{mydarkred}{rgb}{0.64,0,0}
  \hypersetup{ %
    pdftitle={},
    pdfauthor={},
    pdfkeywords={},
    pdfborder=0 0 0,
    pdfpagemode=UseNone,
    colorlinks=true,
    linkcolor=mydarkblue,
    citecolor=mydarkred,
    filecolor=cyan,
    urlcolor=magenta,
    pdfview=FitH}

\newcommand{\tp}{^{\top}}

\DeclarePairedDelimiterX{\infdivx}[2]{[}{]}{%
  #1\;\delimsize\|\;#2%
}

\DeclareMathOperator{\bernoullipdf}{Bernoulli}
\DeclareMathOperator{\normalpdf}{N}
\DeclareMathOperator{\gammapdf}{Gamma}

\usepackage{todonotes}

\title{User Modelling for Avoiding Overfitting in Interactive Knowledge Elicitation for Prediction \footnote{This is the pre-print version. The paper is published in the proceedings of \textit{IUI 2018} conference. Definitive version DOI: \href{https://doi.org/10.1145/3172944.3172989}{https://doi.org/10.1145/3172944.3172989}.}}

\author{Pedram Daee$^\dagger$} 
\author{Tomi Peltola$^\dagger$}
\author{Aki Vehtari}  
\author{Samuel Kaski}

\affil{ Helsinki Institute for Information Technology HIIT, \\ Department of Computer Science, Aalto University \\   \texttt{firstname.lastname@aalto.fi}\\\smallskip  {\small  $^\dagger$Authors contributed equally.} }
\date{}
\begin{document}

\maketitle
\begin{abstract}

  In human-in-the-loop machine learning, the user provides information beyond that in the training data. Many algorithms and user interfaces have been designed to optimize and facilitate this human--machine interaction; however, fewer studies have addressed the potential defects the designs can cause. Effective interaction often requires exposing the user to the training data or its statistics. The design of the system is then critical, as this can lead to double use of data and overfitting, if the user reinforces noisy patterns in the data. We propose a user modelling methodology, by assuming simple rational behaviour, to correct the problem. We show, in a user study with 48 participants, that the method improves predictive performance in a sparse linear regression sentiment analysis task, where graded user knowledge on feature relevance is elicited. We believe that the key idea of inferring user knowledge with probabilistic user models has general applicability in guarding against overfitting and improving interactive machine learning.

\end{abstract}

\section{INTRODUCTION}

\begin{figure*}[t]
\centering
  \includegraphics[width=0.6 \columnwidth,keepaspectratio]{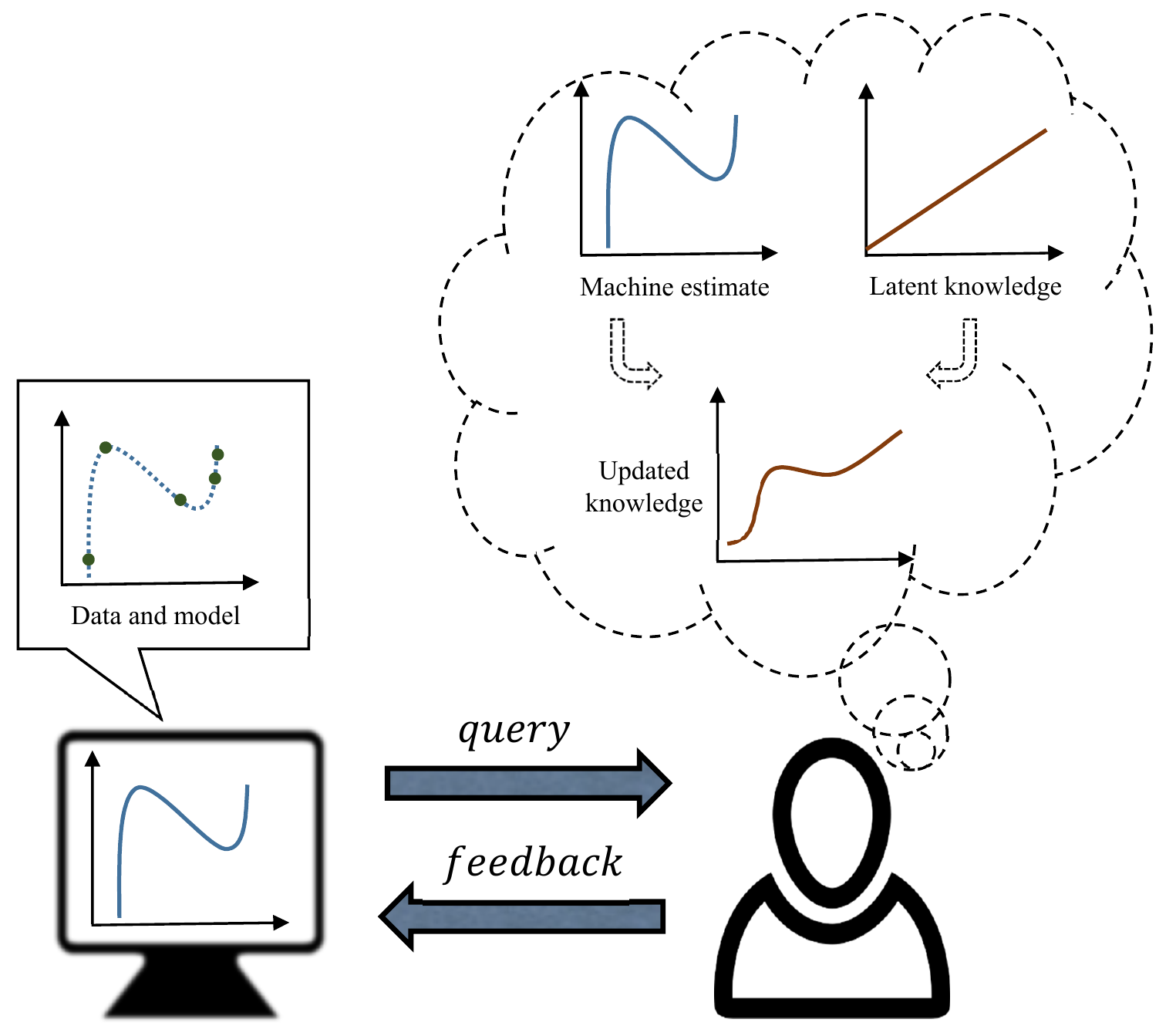}
  \caption{A schematic example of overfitting in human-in-the-loop machine learning. The machine and user are collaborating to improve a regression problem. The machine fits a regression model to the training data (green dots) and employs it to interact with the user (here by visualizing the trained model). Through the interaction, the user learns aspects of the training data and considers them to update her latent knowledge and to form her feedback. This creates a dependency between the feedback and training data that needs to be accounted for in the model to avoid double use of data and overfitting.
  }~\label{fig:interaction_schematic}
\end{figure*}

Interactive and human-in-the-loop machine learning (and the related field of visual analytics) exploits the complementary knowledge and skills of humans and machines to improve performance over automatic training-data-based machine learning and to extend the reach of machine learning systems \cite{Afrabandpey2017, amershi2014power, Daee2017, Holzinger2016, Micallef2017, sacha2017you}. However, the study of how to optimally combine the strengths of humans and machines is still in its early phase, and many possible issues arising from the interaction have not been thoroughly considered yet.

Overfitting, that is, the model fitting to idiosyncrasies in the training data and hence not generalizing to new data, is a thoroughly studied topic in automatic machine learning. Subtle issues with regard to overfitting can arise when introducing a human into the loop. Yet the risk of overfitting seems little discussed in the interactive machine learning literature, although many methods combine user interaction with training-data-based machine learning models. For example, many methods visualize statistics of training data or machine output directly for the user \cite{Afrabandpey2017,Kapoor2010,krause2014infuse,Micallef2017,muhlbacher2013partition,Talbot2009,van2011baobabview} or use the training data to select informative queries to present for the user (e.g., active learning) \cite{Daee2017, settles2010active,sundin2017improving}. Some methods use \textit{validation} datasets in addition to the training set to evaluate the performance. This can help to alleviate overfitting, but if the user can interact with the model based on the results on the validation, the validation set is effectively only another training dataset. A parallel line of research into controlling for overfitting has spawned in adaptive data analysis \cite{dwork2015reusable, russo2016controlling}, which has a related but different goal from interactive machine learning, of exploring the data for interesting hypotheses.

In interactive machine learning, we note that, in particular, the following three steps can lead to overfitting and hence decrease in the performance of the model, as they violate the assumption of independence of the feedback and the training data: (1) showing the training data or some of its statistics to the user, (2) querying the user for feedback, and (3) inputting the feedback back to the model as independent data (a common assumption in machine learning models). Figure~\ref{fig:interaction_schematic} illustrates the induced user behaviour producing a dependency between the feedback and training data. Overfitting can also happen if the user controls some preferences (e.g., cost function weighting or regularization) in the model based on the training or validation data, since the effected improvement in the fit to the training or validation data will not necessarily generalize. The more freedom the user has (or, in this sense, the richer the feedback is), the more problematic this can be.

Rather than tying the hands of the user, improved performance can be attained by accounting for the risk of overfitting in the design of the interactive machine learning system. We propose a user modelling approach in probabilistic models to infer the user's knowledge that is complementary to the training data. We assume that the user behaves rationally, in a simple Bayesian sense \cite{gershman2015computational}, in combining her latent knowledge with the information that the machine reveals of the training data. Given the observed user feedback, we then invert the process to infer the latent user knowledge and use it to update the model. We illustrate the approach in a proof-of-concept user study in a sparse linear regression prediction task, where graded feature relevance knowledge is elicited from the users.

\section{METHOD} \label{method}

\subsection{Overview}

We consider human--machine interaction in probabilistic models. In particular, we consider a situation, where the machine estimates a probabilistic model from training data, and then asks the user for feedback for the learned model. The feedback is included as further data in the model. This procedure can be iterated, although in the specific implementation here, we only consider a single round.

The setup is as follows. The machine has a probabilistic model for a prediction task and a set of training data that it uses to fit the parameters of the model. The machine assumes that the user has knowledge about some aspect of the task and elicits user knowledge by displaying the result of learning from the training data and asking for feedback. This kind of interaction is common in interactive machine learning. If the system naively uses this user feedback to update the model, for example, including it as an observation that is assumed independent of the training data, it risks double use of the data and overfitting, because such assumption would clearly be invalid. On the other hand, building a feedback model that would adequately describe the dependency of the feedback on the training data can be difficult.

We instead propose to infer the latent (unobserved) user knowledge, representing information that the user has beyond that of the training data, from the observed user feedback. This inferred knowledge can then be used to update the machine's model without double use of the training data. To make this feasible, we assume that the user behaves rationally, using the Bayes theorem, in integrating the information sources (her knowledge beyond training data and the information the machine provided). We then invert the process to infer the latent user knowledge.

\subsection{General Mathematical Formulation}

Let the observation model of a training dataset $\mathcal{D}$ be $p(\mathcal{D} \mid \bm{\theta})$, where $\bm{\theta}$ is a vector containing the model parameters, and $p(\bm{\theta})$ be their prior distribution. Given the model and the training dataset, the machine computes the posterior distribution $p(\bm{\theta} \mid \mathcal{D})$ of the parameters using the Bayes theorem. We assume the user has knowledge about some parameter or statistic $\phi$, which is an element in $\bm{\theta}$ or, more generally, a function of $\bm{\theta}$. The machine provides the user with information on $\phi$, for example, its posterior distribution $p(\phi \mid \mathcal{D})$ and asks the user to provide feedback on it.

Let $f$ be the latent (unobserved) user knowledge. Our goal is to infer  from the observed feedback a latent feedback likelihood function $p(f \mid \phi)$ that can be used to update the model to $p(\bm{\theta} \mid \mathcal{D}, f)$. By the assumption of a rational user, the observed feedback is based on the posterior distribution
\begin{equation}
p(\phi \mid \mathcal{D}, f) = \frac{p(f \mid \phi) p(\phi \mid \mathcal{D})}{p(f \mid \mathcal{D})}, \label{eqn:user_modelling}
\end{equation}
where $p(f \mid \mathcal{D}) = \int p(f \mid \phi) p(\phi \mid \mathcal{D}) d\phi$ is the normalization constant.
The technical details on how to invert this to learn $p(f \mid \phi)$ are case-specific. In the next section, we will show how to do this in eliciting feature relevance for sparse linear regression.

\subsection{Feature Relevance Elicitation in Sparse Linear Regression}

We apply the approach to infer user knowledge on feature relevance in sparse linear regression. We use a probabilistic sparse linear regression model described in \cite{Daee2017}, which formulates a linear model to predict the target variable $y$ given a vector of features $\bm{x}$ and uses a spike-and-slab prior to model whether features are included or excluded from the regression:
\begin{align*}
 y_i &\sim \normalpdf(\bm{x}_i\tp \bm{w}, \sigma^2),\\
 \sigma^{-2} &\sim \gammapdf(\alpha_\sigma, \beta_\sigma),\\
 w_j &\sim \gamma_j \normalpdf(0, \tau^2) + (1 - \gamma_j) \delta_0,\\
 \gamma_j &\sim \bernoullipdf(\rho),
\end{align*}
where $i=1,\ldots,N$ runs over $N$ training samples $(y_i, \bm{x}_i) \in \mathcal{D}$, $j=1,\ldots,M$ runs over $M$ features, $\sigma^2$ is a residual variance parameter, the $w_j$ are regression weights, and the $\gamma_j$ are binary variables indicating whether feature $j$ is included in the regression ($\gamma_j = 1$: $w_j$ is a priori normally distributed with variance $\tau^2$) or excluded ($\gamma_j = 0$: $w_j = 0$ via the point mass $\delta_0$). The parameter $\rho$ is the prior expected proportion of included variables. The probabilistic parameters of the model are $\bm{\theta} = (\bm{w}, \sigma^2, \bm{\gamma})$; here, $\alpha_\sigma$, $\beta_\sigma$, $\tau^2$, and $\rho$ are fixed hyperparameters.

To elicit feedback on feature relevance, we show the marginal posterior probability $p(\gamma_j = 1 \mid \mathcal{D})$ to the user and ask her to provide as feedback her estimate of the probability of the feature being relevant for the prediction. To infer the latent user knowledge, we assume that the user's feedback is the posterior probability
\begin{equation}
p(\gamma_j = 1 \mid \mathcal{D}, f_j) = \frac{p(f_j \mid \gamma_j = 1) p(\gamma_j = 1 \mid \mathcal{D})}{Z},\label{eqn:user_modelling_case}
\end{equation}
where $Z$ is the normalization constant
and the latent feedback likelihood is
\begin{equation*}
p(f_j \mid \gamma_j) = A_{f_j} \gamma_j + B_{f_j} (1 - \gamma_j),
\end{equation*}
where $A_{f_j}$ is the likelihood for the latent $f_j$ when $\gamma_j = 1$ and $B_{f_j}$ when $\gamma_j = 0$. Without loss of generality (for using the likelihoods for updating the model later), we can set $A_{f_j} + B_{f_j} = 1$, with $A_{f_j} \in (0, 1)$ and $B_{f_j} \in (0, 1)$.

We infer $A_{f_j}$ (and, consequently, $B_{f_j} = 1 - A_{f_j}$) by solving from the Bayes theorem in Equation~\ref{eqn:user_modelling_case}:
\begin{equation*}
A_{f_j} \propto \frac{p(\gamma_j = 1 \mid \mathcal{D}, f_j)}{p(\gamma_j = 1 \mid \mathcal{D})},
\end{equation*}
where the numerator is the observed feedback given by the user and the denominator is the machine's posterior probability that was shown to the user.

Given a set of observed feedbacks $\mathcal{F}$ from the user, we update the model to $p(\bm{\theta} \mid \mathcal{D}, \mathcal{F})$, where $\bm{\theta}$ denotes all parameters of the model, by updating $p(\bm{\theta} \mid \mathcal{D})$ using the Bayes theorem and the inferred likelihood functions $p(f_j \mid \gamma_j)$ for each feature $j$ with observed feedback.

Computation of the posterior distribution is intractable. Expectation propagation is used to approximate it (see \cite{Daee2017}).

\section{EXPERIMENT AND RESULTS}

\subsection{Sentiment Analysis Task}

We considered the problem of rating prediction from user reviews on Amazon kitchen products. The task and data were previously studied in ~\cite{Daee2017,Hernández-Lobato2015}. The data consist of review texts, represented as bag-of-words with 824 distinct unigram and bigram keywords, and their corresponding 1--5 star ratings. The machine learning system aims to predict the ratings of new reviews (test data), given some training data and external expert knowledge. To make the prediction challenging, the data set was randomly partitioned in 500 training data and 4649 test data (the number of training data is smaller than the number of dimensions). 

The goal of the experiment was introduced to the participants as eliciting domain knowledge from people to help in designing better predictors of product ratings. However, the real research question was to investigate the efficiency of the user model and user interaction in different scenarios. We listed 70 keywords from the reviews and asked the participants to judge the probability of relevance of each keyword in predicting the product rating. The participants could provide feedback by adjusting the probability value between 0 (not-relevant at all) to 1 (absolutely relevant) by moving a slider. The feedback was only recorded after the slider was moved. The participants could skip giving feedback to keywords that they were very uncertain about. The task description provided example keywords \textit{must buy}, \textit{disaster}, and \textit{is} and explained that the first two provide useful information about the product rating, and therefore, they are relevant while the latter is uninformative for rating prediction.

Two systems were implemented, a baseline system where the initial position of sliders were set to the default value 0, and an interactive elicitation (IE) system where the initial positions were set by the machine to the posterior inclusion probabilities $p(\gamma_j=1 \mid \mathcal{D})$ based on the training data. The participants were informed about the initialization method. Both systems use the prediction model and the feedback likelihoods introduced in the previous section with the difference that IE can infer the latent user knowledge based on the proposed user model while the baseline directly applies the user feedback in the model. Following \cite{Daee2017}, the model hyperparameters were set as $\alpha_\sigma = 1,~\beta_\sigma = 1,~\rho = 0.3,$ and $\tau^2=0.01$. Mean squared error (MSE) on test data is used as the performance measure. The code, data, and experiment forms can be found in \href{https://github.com/HIIT/human-overfitting-in-IML}{https://github.com/HIIT/human-overfitting-in-IML}.

\subsection{Results}

48 university students and researchers participated in the user study, 3 were excluded since they either left more than $2/3$ of the questions unanswered or they finished the study in less than 3 minutes (it was unrealistic to go through the form in less than 3 minutes). 23 participants used the baseline and the remaining 22 the IE system.  

\begin{table}
  \centering
  \begin{tabular}{l| r| r| r}
    &  \multicolumn{3}{c}{\small{\textbf{Test MSE $\pm$ STD }}} \\
    {\small\textit{Condition}}
    & {\small \textit{No feedback}}
    & {\small \textit{User feedback}}
    & {\small \textit{User model}} \\
    \hline
    Baseline & $1.835$ & $1.749 \pm 0.050$ & NA \\
    IE & $1.835$ & $1.744 \pm 0.045$ & $1.705 \pm 0.038$ \\
  \end{tabular}
  \caption{Mean and standard deviation (STD) of MSE on test data for the two systems in different conditions.}~\label{tab:MSE}
\end{table}

\begin{figure}
\centering
  \includegraphics[width=0.6\columnwidth]{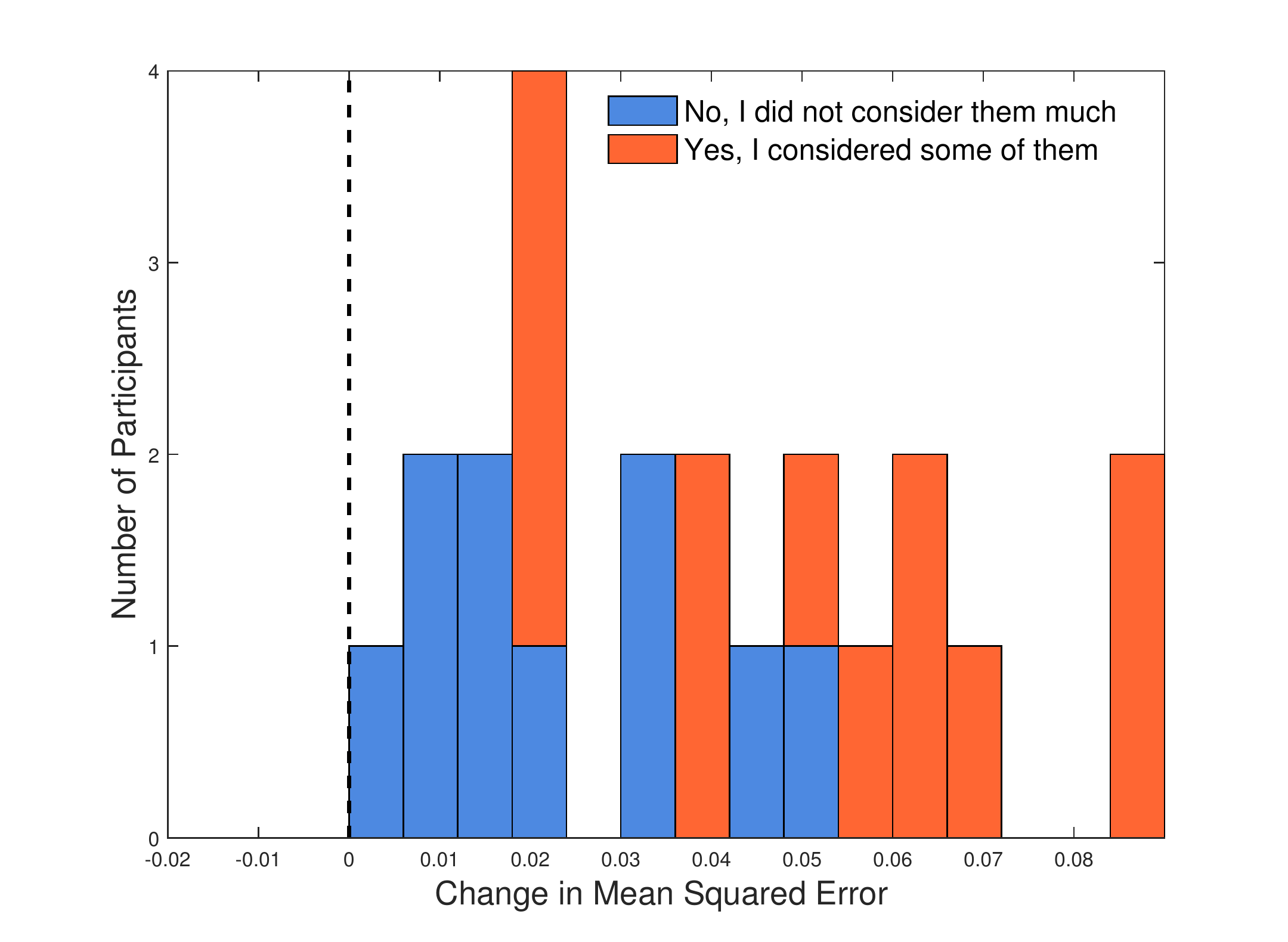}
  \caption{Stacked histogram of MSE change (the difference between MSE after directly using the feedbacks and MSE after the correction done by the propose user model) for 22 participants in IE system. The participants are grouped based on their answer to the question on whether they found the machine estimates useful or not. User modelling has improved MSE values for all participants.}~\label{fig:MSE_improvement}
\end{figure}

Table~\ref{tab:MSE} shows the average test MSE for participants of the two systems before and after receiving feedback and also after the correction done by the proposed user model in IE. The feedback improved the predictive performance in both systems (p-value$=3\times10^{-8}$ in baseline and p-value$=5\times10^{-9}$ in IE without user model, using paired-sample \textit{t}-test). This shows, as analogously claimed in \cite{Daee2017}, that the participants, on average, have the necessary knowledge to improve the prediction. More prominently, the predictive performance further improved in the IE system after inferring the latent user knowledge using the user model (p-value$=7\times10^{-7}$ in paired-sample \textit{t}-test between directly using the feedbacks and after employing the user model). Moreover, the user model improved the predictions for each individual user (Figure~\ref{fig:MSE_improvement}). In a post-questionnaire, we asked the participants of the IE system about the usefulness of the machine estimates: 12 participants answered that they considered them when giving some of the answers and the remaining 10 responded that they did not consider machine estimates that much. On average, the predictions of the former group improved more with the user model (Figure~\ref{fig:MSE_improvement}).

The feedback the participants gave to the keywords differed between the two systems in several aspects. Table~\ref{tab:words_smallp} lists average feedback values and machine estimates for 10 keywords with the lowest nominal p-values (two-sample t-test without assuming equal variances between the two systems), showing the keywords that were the most different between the systems. The average correlation to machine estimates for users in IE system was 0.46 (0.33 for the baseline system) and the average variance of feedbacks on keywords was 0.043 (0.060 for the baseline system). These support the hypothesis that the participants considered the machine estimates in the IE system. In the IE system, the average correlation to machine estimates after the correction done by the user model declined to -0.017 which suggests that the user model was successful in reducing the dependency to the training data.

\begin{table}
\centering
\begin{tabular}{c|c|c|c|c}
        {\small \textit{\textbf{Keyword}}} & {\small \textit{\textbf{Machine}}} & {\small \textit{\textbf{IE}}}  & {\small \textit{\textbf{Baseline}}} & {\small \textit{\textbf{P-value}}} \\
        \hline
best & 0.89     & 0.90      & 0.80      & 0.014     \\
great & 1.00     & 0.95      & 0.83      & 0.031     \\
disappointed & 0.99     & 0.96      & 0.85      & 0.038     \\
heavy & 0.20     & 0.36      & 0.52      & 0.038     \\
buy this & 0.70     & 0.77      & 0.63      & 0.053     \\
steel & 0.19     & 0.12      & 0.24      & 0.073     \\
line & 0.34     & 0.15      & 0.06      & 0.088     \\
don't & 0.47     & 0.54      & 0.38      & 0.090     \\
recommend & 0.16     & 0.74      & 0.84      & 0.102    \\
good & 0.26     & 0.71     & 0.81      & 0.115  
\end{tabular}
\caption{Difference between user feedback in the baseline and IE systems (without user model).}~\label{tab:words_smallp}
\end{table}

\section{DISCUSSION AND CONCLUSION}

We described a user modelling methodology in probabilistic models for disentangling latent user knowledge from observed user feedback that was given in response to machine revealing information from the training data. We used this to guard against double use of the training data and overfitting in interactive machine learning. The proof-of-concept user study in interactive knowledge elicitation of feature relevance information in sparse linear regression showed the potential of the approach for improving prediction performance.

Our approach is based on a simple rationality assumption of the user. It is, however, unreasonable to assume that users would in general behave completely rationally. In particular, the amount information and the way it is presented to the user are important factors that should be considered in designing interactive machine learning systems and user models. In our experiment, the interaction was based on reporting probability values through sliders. Such probability elicitation can be prone to the anchoring effect \cite{Garthwaite2005, Tversky1974}. Arguably, similar psychological mechanisms will appear in many interactive machine learning applications, since the machine often needs to guide the user for efficient interaction. This paper demonstrates that as long as the user modelling is able to capture the main sources of defects in the interaction, it would be expected to improve the results.

We believe that, as the field of interactive machine learning matures, more consideration will be put into the possible defects in the human--machine interaction. In particular, better user modelling will allow the user to behave naturally in providing feedback to the machine, while the machine (that is, the design of the machine learning system and models) will account for the inevitable human factors and biases \cite{amershi2014power, Garthwaite2005, NIPS2016_6155, Kulesza2014, Newell2016, sacha2017you, Tversky1974} to optimally combine the complementary knowledge and skills of the user and the machine. Our work is a step towards this with regard to overfitting in the interaction.

\section{ACKNOWLEDGMENTS}
This work was financially supported by the Academy of Finland (Finnish Center of Excellence in Computational Inference Research COIN; Grants 295503, 294238, 292334, and 284642), Re:Know funded by TEKES. We acknowledge the computational resources provided by the Aalto Science-IT Project. We thank Marta Soare for collaboration and helpful comments in early stage of the project.






\end{document}